\documentclass[11pt]{article}
\usepackage{graphicx}

\newsavebox{\hflrar}
\sbox{\hflrar}{\makebox[0pt][l]
{${\scriptstyle \leftharpoonup}$}{${\scriptstyle \rightharpoonup}$}}

\def \to {\rightarrow}

\begin{document}
\begin{center}
{\Large\bf Resummation of Large Logarithms in $\gamma^* \pi^0 \to \gamma $}
\vskip 10mm
F. Feng$^1$, J.P. Ma$^1$ and Q. Wang$^{2}$    \\
{\small {\it $^1$ Institute of Theoretical Physics, Academia Sinica,
Beijing 100080, China }} \\
{\small {\it $^2$ Department of Physics and Institute of Theoretical
Physics, Nanjing Normal University, Nanjing, Jiangsu 210097, P.R.China}} \\
\end{center}

\vskip 1cm
\begin{abstract}
In the collinear factorization of the form factor for the transition
$\gamma^* \pi^0 \to \gamma$  the hard part contains double log terms as $\ln^2 x$
with $x$ as the momentum fraction of partons from $0$ to $1$.
A simple exponentiation for resummation leads to divergent
results. We study the resummation of these $\ln^2 x$ terms. We show that
the $\ln^2 x$ terms come partly from the light-cone wave function(LCWF)
and partly from the form factor. We introduce a jet factor to factorize
the $\ln^2 x$ term in the form factor.
To handel the $\ln^2 x$ terms from the LCWF we introduce
a nonstandard light-cone wave function(NLCWF) with
the gauge links off the light-cone direction. An interesting relation
between two wave function is found.
With the introduced
NLCWF and the jet factor we can re-factorize the form factor and
obtain a new hard part which does not contain terms with $\ln^2 x$.
Beside the renormalization scale $\mu$ the introduce NLCWF and jet factor
have extra scales to characterize their $x$-behaviors. Using the evolutions
of the extra scales and the relation we can do the resummation
perturbatively in sense that the LCWF is the only nonpertubative object
in the resumed formula. Our results with some models of LCWF
show that there is a significant difference between numerical predictions with
the resummation and that without the resummation, and the resummed predictions
can describe the experimental data.

\vskip 5mm \noindent
\end{abstract}
\vskip 1cm
\par\vfil
\eject

\noindent
{\bf\large 1. Introduction}
\par\vskip15pt
Predictions with perturbative  QCD  for an exclusive process can be made
if it contains short-distance effects beside long-distance effects.
In order to use the perturbative theory of QCD one needs
to separate or factorize long-distance- and short-distance effects. Only the latter,
which are characterized by a large energy scale denoted generically as $Q$,
can be studied with perturbative QCD.
It has been proposed long time ago that
such a process can be studied by an expansion of
the amplitude in $1/Q$, corresponding to an expansion of QCD
operators in twist\cite{BL,CZrep}. The leading term can be
factorized as a convolution of a hard part and light-cone wave
functions of hadrons. The light-cone wave functions are defined with
QCD operators, and the hard part describes hard scattering of
partons at short distances. The hard part can be safely calculated
with perturbative QCD in the sense that it does not contain any
infrared(I.R.)- and collinear divergences.
This is the so-called collinear
factorization. In this factorization the transverse
momenta of partons in parent hadrons are also expanded in the hard
scattering part and they are neglected at leading twist.
\par
Although the hard part does not contain any of I.R- and collinear singularities,
but as an perturbative expansion it contains large logarithms at higher orders of
$\alpha_s$. These large logarithms are dangerous and can spoil the expansion
in the sense that the expansion does not convergent. A resummation
of large logarithms
is often needed to have a reliable prediction. In this paper we study
the resummation in the process $\gamma^* \pi^0 \to \gamma$.
Theoretically, the process
has been studied with QCD
factorization extensively\cite{COF1,COF2,KR,GHM,MR,C1L,C2L}.
Experimentally it has been studied too\cite{EXP}.
With the collinear factorization the form factor $F(Q)$ characterizing the process
can be written as
\begin{equation}
   F(Q) \sim \phi \otimes H \left \{ 1 + \frac{\Lambda^2}{Q^2} \right\}.
\end{equation}
In the above  $\phi$ is the light-cone wave function(LCWF) of $\pi^0$,
$H$ is the hard part, and $Q^2$ is the virtuality
of the virtual photon. The correction to the factorized form is power-suppressed and is proportional
to $\Lambda^2/Q^2$.
$\Lambda$ is any possible soft scale characterizing nonperturbative
scale, like $\Lambda_{QCD}$, the mass of $\pi$ etc.. This scale
is about several hundreds MeV.
In \cite{C1L} the hard part in the collinear factorization
has been calculated at one-loop, its two-loop result can be found in
\cite{C2L}. At one-loop level one finds that $H$ contains a double log $\alpha_s\ln^2 x$,
where $x$ is the momentum fraction carried by a parton with $0\leq x \leq 1$.
It is expected that at $n$-loop level, it will contain $ \alpha_s^n \ln^{2n} x$.
Those double log terms will become divergent when $x$ is approaching to zero
and can spoil the perturbative
expansion of $H$. Although the double log terms are integrable with $x$ in the convolution
and give finite contributions, but they are significant corrections.
The purpose of our study is to resum those double log terms.
\par
In order to resum those double log terms we need to understand their
origin. Given the factorized form in the above, $H$ will receive
contributions from the form factor and the LCWF. When we use a
finite quark mass to regularize the collinear singularities, we can
show that the log terms come from the form factor and the LCWF.
These terms can be re-factorized by introducing  a jet factor and a
nonstandard light-cone wave function(NLCWF), where the jet factor
contains the double log from the form factor and the NLCWF doe not
have double log. In introducing these two objects in the
re-factorization, two extra scales, which will be explained later,
are introduced. One of them is related to the $\ln^2 x$ term from LCWF,
while another is related to $\ln^2 x$ in the form factor.
The new hard part $\tilde H$ from the
re-factorization will not contain  $\ln^2 x$.
With evolution equations of these scales we are able to resum the
$\ln^2 x$ terms. Our approach is similar to the threshold resummation in
inclusive processes studied in \cite{Ster}. There exists an interesting
relation between the two wave functions.
In our approach, after
the resummation of $\ln^2 x$, only the LCWF appears in the form
factor as a nonperturbative object, other quantities can be
calculated with perturbative QCD. With the knowledge of the LCWF we
are able to give numerical results and to make a comparison with
experiment.
\par
It should be noted that it is possible to resum the large log terms in exclusive processes
by taking transverse momenta of partons into account\cite{BS,LS}. By introducing
transverse-momentum-dependent(TMD) light-cone wave-functions one can make
TMD of $k_t$ factorization in terms of the TMD light-cone wave functions.
However, the consistence of the factorization needs to be carefully checked
beyond tree-level. Recently, this has been studied in a series of papers\cite{TMD1,TMDB}, where
a consistent definition of TMD light-cone wave functions is proposed and
the TMD factorization is examined beyond tree-level for some simple cases.
However, unlike LCWF's there is little knowledge about TMD light-cone wave functions.
Therefore, it may be difficult to give detailed predictions and to compare
with experiment.
\par
Our paper is organized as the following: In Sec. 2 we introduce our notations
and explain the origin of $\ln^2 x$. In Sec. 3 we introduce NLCWF and present
a one-loop study of the NLCWF. We also show that there is a perturbative
relation between LCWF and NLCWF. In Sec. 4 we introduce our jet factor and derive
the factorization formula with the jet factor and NLCWF. In Sec. 5 we give
our resummation formula. We present our numerical results in Sec.6, where
a comparison with experiment is also given. Sec.7 is our conclusion.

\par\vskip20pt
\noindent
{\bf\large 2. Notations and the Origin of $\ln^2 x$}
\par\vskip15pt
We consider the process:
\begin{equation}
\pi^0 +\gamma^*\to  \gamma
\end{equation}
where $\pi^0$ carries the momentum $P$ and the real photon momentum
$p$. We use the  light-cone coordinate system, in which a
vector $a^\mu$ is expressed as $a^\mu = (a^+, a^-, \vec a_\perp) =
((a^0+a^3)/\sqrt{2}, (a^0-a^3)/\sqrt{2}, a^1, a^2)$ and $a_\perp^2
=(a^1)^2+(a^2)^2$. We take a frame in which the momentum $P$ and $p$ are:
\begin{equation}
P^\mu =(P^+, P^-,0,0), \ \ \ \ \ \  p^\mu =(0,p^-,0,0).
\end{equation}
We will consider the case that the virtual photon has the large
negative virtuality $q^2 =(P-p)^2=-2P^+ p^-=-Q^2$. The process can
be described by matrix element $\langle \gamma (p, \epsilon^*)\vert
J^\mu_{e.m.}\vert \pi^0(P)\rangle$, which is
parameterized with the form factor $F(Q^2)$:
\begin{equation}
\langle \gamma (p, \epsilon^*)\vert J^\mu_{e.m.}\vert \pi^0(P)\rangle
= i e^2 \varepsilon^{\mu\nu\rho\sigma} \epsilon^*_{\nu} P_{\rho} p_{\sigma}
F(Q^2).
\end{equation}
$e$ is the charge of proton, i.e., $\alpha =e^2/(4\pi)$.
\par
In the collinear factorization the form factor can be factorized as
\begin{eqnarray}
  F(Q^2) = \frac { Q_u^2 - Q_d^2}{\sqrt{2}}\frac{1}{ Q^2} \int_0^1 dx \phi (x,\mu) H (x,Q,\mu) \left [ 1
     + {\mathcal O} ( \frac{\Lambda^2}{ Q^2} ) \right ],
\end{eqnarray}
where $\phi$ is the LCWF of $\pi^0$, $H$ is a perturbative function or a hard part.
$Q_u$ and $Q_d$ are the electric charge fraction of $u$ and $d$ quark in unit of $e$,
respectively. $\phi$ is defined with QCD operators:
\begin{equation}
 \phi(x, \mu) = \int \frac{ d z^- }{2\pi}
    e^{ik^+z^- }
 \langle 0 \vert \bar q(0) L_n^\dagger (\infty, 0)
  \gamma^+ \gamma_5 L_n (\infty,z^- n) q(z^- n ) \vert \pi^0(P) \rangle,
\end{equation}
where the gauge link is defined along the light-cone direction $n^\mu=(0,1,0,0)$ as:
\begin{equation}
L_n (\infty, z) = P \exp \left ( -i g_s \int_{0} ^{\infty} d\lambda
     n \cdot G (\lambda n+ z ) \right ) .
\end{equation}
\par
To find the hard part $H$, we replace the
hadronic state with the partonic state:
\begin{eqnarray}
&& \vert \pi^0 (P) \rangle \to \vert q(k_q), \bar q(k_{\bar q})\rangle,
\ \ \ \ k_q^\mu =(k_q^+, k_q^-,0,0), \ \ \ \  k_{\bar q}^\mu
=(k_{\bar q}^+, k_{\bar q}^-, 0,0 )
\nonumber\\
&&
k^2_q = k^2_{\bar q} =m^2, \ \ \  k_q^+ = x_0 P^+, \ \ \ \  k_{\bar q}^+ = (1-x_0) P^+ = \bar x_0 P^+,
\end{eqnarray}
where we use a small but finite quark mass $m$ to regularize collinear singularities.
The form factor calculated with the partonic state will in general
contain collinear singularities. The LCWF calculated with the partonic state
will also have collinear singularities. If the collinear factorization holds,
the singularities of the form factor and the LCWF will be the same so that
the hard part $H$ will not contain any collinear- and I.R. singularities.
Here we examine this explicitly and show the origin of $\ln^2 x$.
\par


\begin{figure}[hbt]
\begin{center}
\includegraphics[width=5cm]{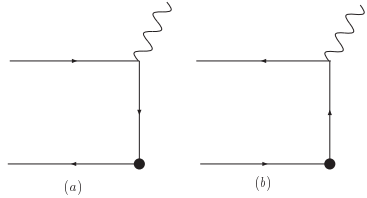}
\end{center}
\caption{Feynman diagrams of tree-level contributions to the
partonic scattering. The black dot denotes the insertion of the
electric current operator corresponding to the virtual photon.}
\label{Feynman-dg1}
\end{figure}
\par
With the partonic state, the LCWF  at tree level reads:
\begin{equation}
\phi^{(0)} (x, \mu) =  \delta (x-x_0) \phi_0, \ \ \ \ \  \phi_0 =\bar
v(k_{\bar q}) \gamma^+ \gamma_5 u(k_q) /P^+.
\end{equation}
At the leading order, the form factor receives contributions from diagrams in Fig.1.
It is straightforward to obtain the tree-level result from Fig.1:
\begin{equation}
F(Q^2)\vert_{1a} = \phi_0 \frac{1}{3\sqrt{2} Q^2 x_0}, \ \ \ \ F(Q^2)\vert_{1b} =
\phi_0 \frac{1}{3\sqrt{2} Q^2(1-x_0)}=\phi_0 \frac{1}{3\sqrt{2} Q^2\bar x_0}.
\end{equation}
We will always use the notation $\bar u =1-u$.
Combining it with the tree-level result of the LCWF, we can
obtain a factorized form for the form factor at tree-level:
\begin{equation}
F(Q^2) = \frac{1}{3\sqrt{2} Q^2} \int dx  \phi (x,\mu) \left [
\frac{1}{x} +\frac{1}{\bar x} \right ], \ \ \ \ \  H^{(0)} (x,Q, \mu) =\frac{1}{x} +\frac{1}{\bar x}.
\end{equation}
\par
At one-loop level,
there are 12 Feynman diagrams, 6 of them are given in Fig.2. The other 6 diagrams are
obtained from those in Fig.2 by reversing the direction of the quark
line, i.e., through charge conjugation. The diagrams in Fig.2 represent the correction to Fig.1a, and
the other 6 diagrams for the correction to Fig.1b. Two corrections
are related each other by charge conjugation. Hence we will need to
study how the contributions from Fig.2 can be factorized.
\par

\begin{figure}[hbt]
\begin{center}
\includegraphics[width=12cm]{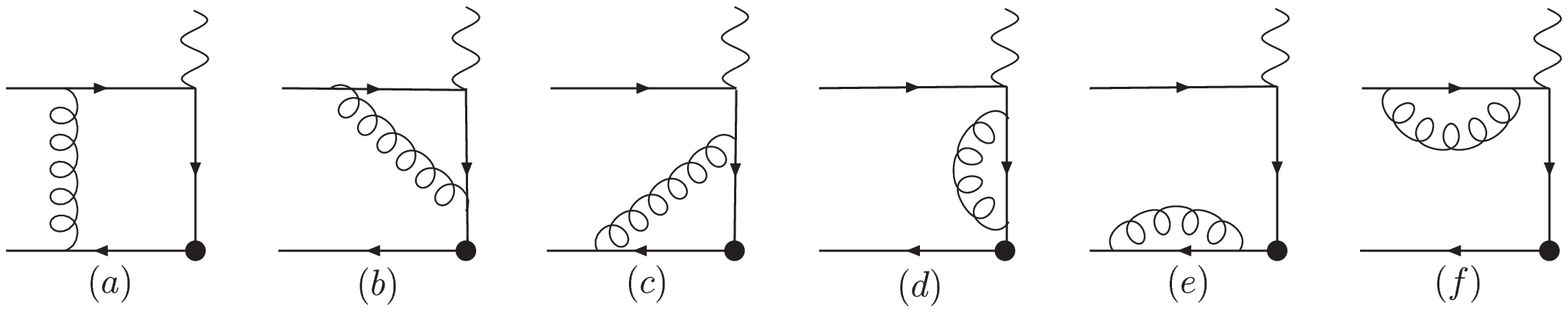}
\end{center}
\caption{Feynman diagrams of the one-loop corrections to Fig.1a.}
\label{Feynman-dg4}
\end{figure}
\par
The contributions except that of Fig.2a can be calculated in a
straightforward way. We give gluons a small mass $\lambda$ to regularize
I.R. singularities.
The results are:
\begin{eqnarray}
F(Q^2)\vert_{2e} &=& F(Q^2)\vert_{2f} = F(Q^2)\vert_{1a} \cdot
\frac{\alpha_s}{6\pi}
     \left [ -\ln\frac{\mu^2}{m ^2} -2 \ln\frac{\lambda^2}{m ^2} -4 \right
],
\nonumber\\
F(Q^2)\vert_{2d} &=& F(Q^2)\vert_{1a}\cdot \frac{-\alpha_s}{3\pi} \left [
\ln\frac{\mu^2}{Q^2} -\ln x_0 +1 \right],
\nonumber\\
F(Q^2)\vert_{2b} &=&F(Q^2)\vert_{1a}\cdot \frac{2 \alpha_s}{3\pi} \left [
\frac{1}{2}\ln\frac{\mu^2}{Q^2} + \ln\frac{Q^2}{m ^2} +\frac{1}{2}\ln x_0
\right],
\nonumber\\
F(Q^2)\vert_{2c} &=& F(Q^2)\vert_{1a}\cdot \frac{2 \alpha_s}{3\pi}
\left \{ \frac{1}{\bar x_0} \left [ -\frac{1}{2} \ln^2 x_0 +2 \ln \bar x_0 \ln x_0+ \ln\frac{Q^2}{m^2}\ln x_0
-\frac{\pi^2}{3} +2 {\rm Li_2}(x_0)  \right ]
\right.
\nonumber\\
   && \left. \ \ \  +\frac{1}{2} \ln\frac{\mu^2}{Q^2} - \ln\frac{m^2}{Q^2}
-\frac{2+x_0}{2\bar x_0} \ln x_0  \right \}.
\label{1LF}
\end{eqnarray}
These results can also be found in \cite{TMD1}.
From the above, the $\ln^2 x$ term comes only from Fig.2c.
The origin of this $\ln^2 x$ is the following: The quark propagator
connecting the vertex which emits the real photon carries the momentum
$(xP^+, -p^-, 0,0) + {\mathcal O}(m^2)$. If $x$ becomes small and goes to zero,
the momentum becomes light-cone-like.
If the momentum of the exchange gluon is in the region collinear
to the $-$-direction, after the loop integration a collinear singularity regularized
by the small $x$ appears. However, the region also overlaps with the infrared region
where all components of the momentum are at order of $xQ$. This region
generates an I.R. singularity which is also regularized by the small $x$.
Therefore, the contribution from Fig.2c contains  $\ln^2 x$, one comes from
the collinear singularity and another from the infrared singularity.
We will show later that the dominant contribution containing $\ln^2 x$ in these regions
can be obtained by the eikonal approximation and
can be factorized by using the method suggested in \cite{CK}.
It has also been suggested by using a jet factor to absorb the $\ln^2 x$ term\cite{HNLI}. 
With a similar analysis one can show that the contribution from Fig. 2b does not
contain the collinear singularity related to the $-$-direction when $x$ becomes
small. It contains only the infrared singularity regularized by $x$. Hence
it does not contain $\ln^2 x$ as shown explicitly in Eq.(12).
\par

\begin{figure}[hbt]
\begin{center}
\includegraphics[width=8cm]{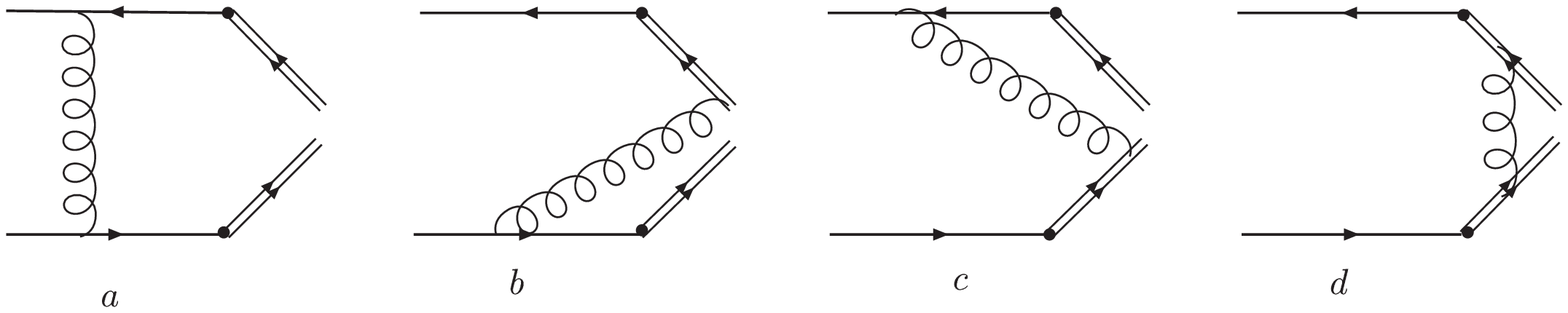}
\end{center}
\caption{Feynman diagrams of the one-loop corrections to LCWF. The double line
 stands for the gauge link}
\label{Feynman-dg4}
\end{figure}
\par
The one-loop correction of the LCWF is given by some of diagrams in Fig.3 and Fig.4.
The diagrams with the gluon-exchange between gauge links give no contribution
here because $n^2=0$.
The one-loop part of LCWF is the sum:
\begin{equation}
\phi ^{(1)} (x,\mu) = \phi (x,\mu)\vert_{3a}+\phi (x,\mu)\vert_{3b} +
\phi(x,\mu) \vert_{3c} +  \phi (x,\mu)\vert_{4a}+ \phi (x,\mu)\vert_{4c}
+  \phi (x,\mu)\vert_{4d}+ \phi (x,\mu)\vert_{4f}  ,
\end{equation}
The one-loop results can be found in \cite{TMD1}. They are:
\begin{eqnarray}
\phi (x,\mu)\vert_{4a} &=& \phi (x,\mu)\vert_{4d} =
 \frac{\alpha_s}{6\pi}
    \left [ -\ln \frac{\mu^2}{m_q^2} +2 \ln\frac{m_q^2}{\lambda^2} -4 \right ] \phi_0,
\nonumber\\
\phi (x,\mu)\vert_{3c} +\phi (x,\mu)\vert_{4f} &=&  -\frac{2\alpha_s}{3\pi} \phi_0
 \theta(x-x_0) \left [ \frac{\bar x}{\bar x_0 (x-x_0)} \ln \frac{m_q^2 (x-x_0)^2}{\mu^2 \bar x_0 ^2} \right]_+,
\nonumber\\
\phi (x,\mu)\vert_{3b} +\phi (x,\mu)\vert_{4c} &=&  \frac{2\alpha_s}{3\pi} \phi_0
 \theta(x_0-x) \left [ \frac{x}{x_0(x-x_0)} \ln \frac{m_q^2 (x-x_0)^2}{\mu^2 x_0^2} \right]_+ .
\end{eqnarray}
\par

\begin{figure}[hbt]
\begin{center}
\includegraphics[width=7cm]{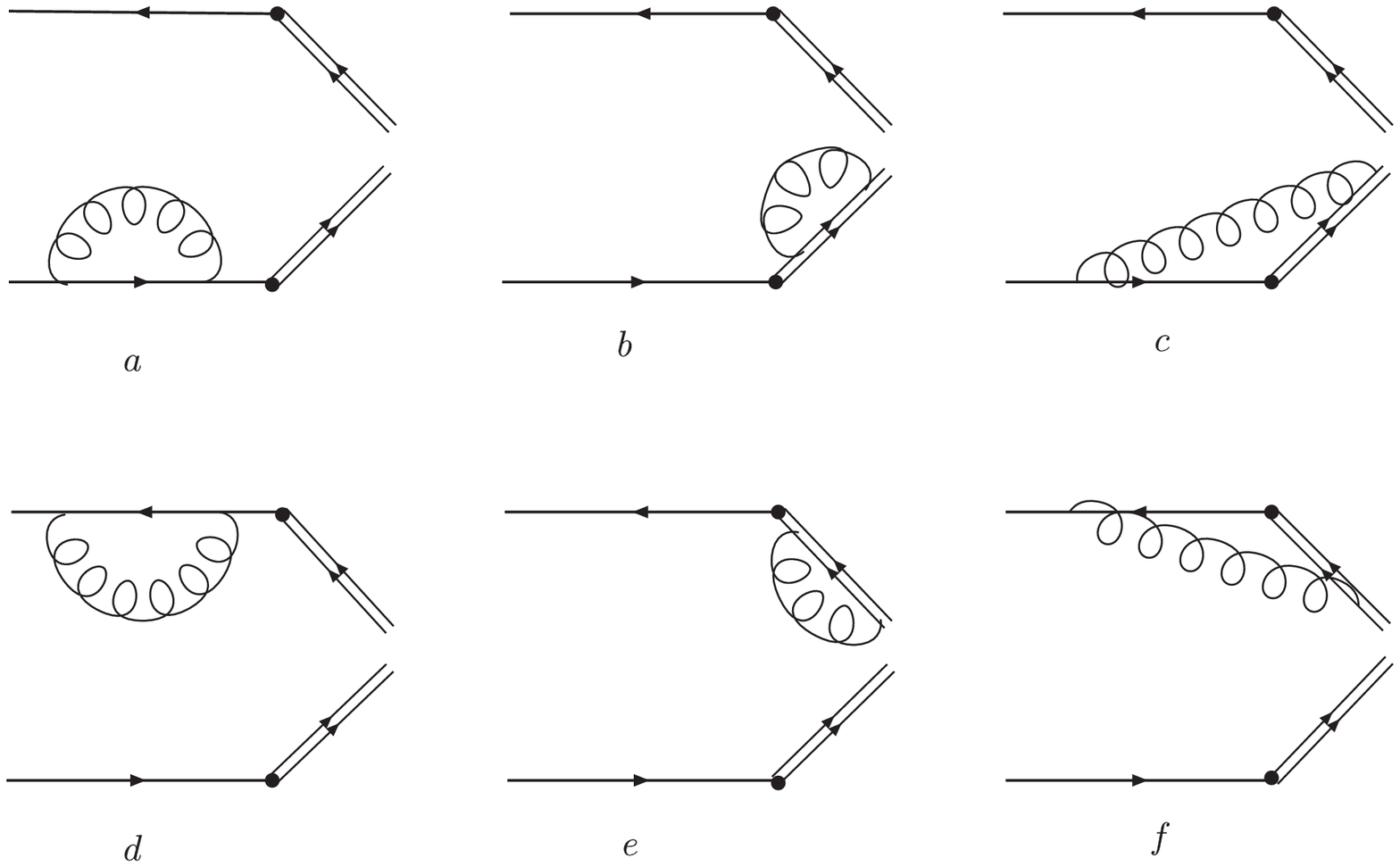}
\end{center}
\caption{Feynman diagrams of the one-loop corrections to LCWF. The double line
 stands for the gauge link}
\label{Feynman-dg4}
\end{figure}
\par
With the above results the one-loop contribution of $H$ can be determined as:
\begin{equation}
  H^{(1)} (x_0, Q,\mu) = F^{(1)} (Q) - \int_0^1 dx \phi ^{(1)} (x,\mu) H^{(0)} (x,Q,\mu).
\end{equation}
It is obvious that the contributions from Fig.2e and Fig.2f are already
contained in the
contribution of Fig.4a and Fig. 4d of the LCWF, respectively.
The contribution from Fig.3a has a complicated
expression. However, for the purpose of the factorization, we only need
the contribution at the leading power of $Q^2$. We have the result for the
combination:
\begin{eqnarray}
 Q^2 F(Q^2)\vert_{2a} -\phi \vert_{3a} \otimes H^{(0)}_{1a}
   = -\frac{2\alpha_s}{3\pi} \frac{1}{\bar x} \ln x
  \left [ \ln\frac{Q^2}{\mu^2} -1 + \frac{1}{2} \ln x\right ]
   +
  {\mathcal O}( Q^{-2}).
\end{eqnarray}
The convolution of other one-loop parts of LCWF reads:
\begin{eqnarray}
\int_0^1 \frac{dx}{x}  \phi (x,\mu)\vert_{3c +4f}
 &=& \frac{2\alpha_s } {3\pi} \frac{1}{x_0 \bar x_0} \left \{ -\ln^2 x_0 +2 \ln \bar x_0 \ln x_0
       + \ln x_0 \ln \frac{\mu^2}{m^2}   +2{\rm Li}_2(x_0)
\right.
\nonumber\\
     && \left.   -\frac{\pi^2}{3}
            + \bar x_0 \ln \frac{\mu^2}{m^2} + 2 \bar x_0 \right \} ,
\nonumber\\
\int_0^1 \frac{dx}{x} \phi (x,\mu)\vert_{3b+4c}
 &=& \frac{2\alpha_s}{3\pi x_0 } \left [ \ln \frac{\mu^2}{m^2} +2 \right ].
\end{eqnarray}
We note that the result in the first line can be used to subtract collinear singularities in Fig.2c
and that in the second line can be used for Fig. 2b. With these results one can extract
the contributions of $H$ from Fig.2c and Fig.2b:
\begin{eqnarray}
H^{(1)} (x, Q,\mu) \vert_{2c} &=& \frac{\alpha_s}{3\pi x} \left \{ \frac{1}{\bar x} \left [ \ln^2 x
    -2 \ln x \ln\frac{\mu^2}{Q^2}  -(2+x) \ln x \right ] -\ln\frac{\mu^2}{Q^2} -4 \right \},
\nonumber\\
H^{(1)} (x, Q,\mu) \vert_{2b} &=& \frac{\alpha_s}{3\pi x}  \left \{ -\ln\frac{\mu^2}{Q^2}
    +\ln x -4 \right \}.
\end{eqnarray}
Finally the one-loop part of $H$ can be given:
\begin{equation}
  H(x,Q,\mu) = \frac{1}{x} +\frac{\alpha_s}{3\pi x} \left [ \ln^2 x -\frac{x}{\bar x} \ln x
        -9  + \ln\frac{Q^2}{\mu^2} \left (3 + 2 \ln x \right ) \right ] + (x\to \bar x) +
        {\mathcal O} (\alpha_s^2)
\end{equation}
\par
From Eq.(17) we can see that the LCWF gives also a contribution with $\ln^2 x$ to $H$.
The origin of this double log is that we use the light-cone gauge link. With the light-cone gauge link
the contribution from Fig.3c and Fig.4f has a light-cone singularity beside a collinear
singularity.
The light-cone singularity is canceled in the sum.
If we use gauge links with non light-cone vectors, the light-cone singularity will be regularized
by the deviation of the vectors from the light-cone vector $n$.
\par
The obtained $H$ behaves like $xH \sim 1 +\alpha_s \ln^2 x/(3\pi)$ when $x$ goes to zero.
A resummation with a simple exponentiation does not work because of the
$+$-sign in the front of the $\ln^2 x$ term. Inspecting the one-loop part $H$
one may chose $\mu$ as $\mu^2 = \sqrt{x} Q^2$ to kill the $\ln^2 x$ term.
However, for small enough $x$ the scale
becomes so small that perturbative QCD can not be used. It seems that
one needs extra nonperturbative objects beside the LCWF
to complete the resummation.
We will show in our work that the resummation can be done without those extra
nonperturbative objects.
\par
Before ending the section we would like to discuss the case if the dimensional
regularization is used to regularize collinear singularities. In this case,
the origin of $\ln^2 x$ is different than that with a quark mass.
However the hard part is the same and it is expected that quantities which are free from collinear
singularities will not depend how the collinear singularities are regularized.
\vskip20pt
\noindent
{\large\bf 3. Nonstandard Light Cone Wave Function}
\par\vskip15pt
As discussed before, one can use non-light cone gauge links to define
nonstandard light cone wave functions. A possible definition
is a straightforward generalization of Eq.(6):
\begin{eqnarray}
 \phi_{+}(x, \zeta, \mu) \sim  \ \int \frac{ d z^- }{2\pi}
    e^{ik^+z^- }
 \langle 0 \vert \bar q(0) L_u^\dagger (\infty, 0)
  \gamma^+ \gamma_5 L_u (\infty,z^-n) q(z^-n) \vert \pi^0(P) \rangle,
\end{eqnarray}
where the gauge link is
\begin{equation}
L_u (\infty, z) = P \exp \left ( -i g_s \int_{0} ^{\infty} d\lambda
     u\cdot G (\lambda u+ z ) \right ),  \ \ \  u^\mu =( u^+, u^-,0,0) .
\end{equation}
This definition is gauge invariant.
The defined NLCWF depends on an extra parameter
\begin{equation}
\zeta^2 = \frac{2 u^- (P^+)^2}{u^+}\approx \frac{ 4 (u\cdot P)^2}{u^2}.
\end{equation}
We will take the limit $u^- >> u^+$ or $\zeta \to \infty$.
The limit $\zeta\to\infty$ should be understood as that we do not take the contributions
proportional to any positive power of $u^+/u^-$ into account.
It has no light-cone singularities as we will show through
our one-loop result.
\par
At tree-level the NLCWF is the same as the LCWF.
At one-loop level, there are contributions given by all diagrams
given in Fig.3 and Fig.4. However, the contributions
from interactions between gauge links will cause
some problems, especially the contribution from Fig.3d.
It should be noted that the contributions from interactions between gauge links,
i.e., those from Fig.3d, Fig.4b and Fig.3e, have no corresponding contributions
in the form factor. A direct calculation shows that the contribution from Fig.3d
is not zero when $x=0$ or $x=1$. When this contribution convoluted
with the tree-level hard part $H^{(0)}$ it will lead to divergences.
Therefore these contributions need to be subtracted and a modification
of the above definition is needed. Since only interactions between gauge links
are involved in these contributions, one can consider
to use products of gauge links to subtract them.
\par
\par
\begin{figure}[hbt]
\begin{center}
\includegraphics[width=8cm]{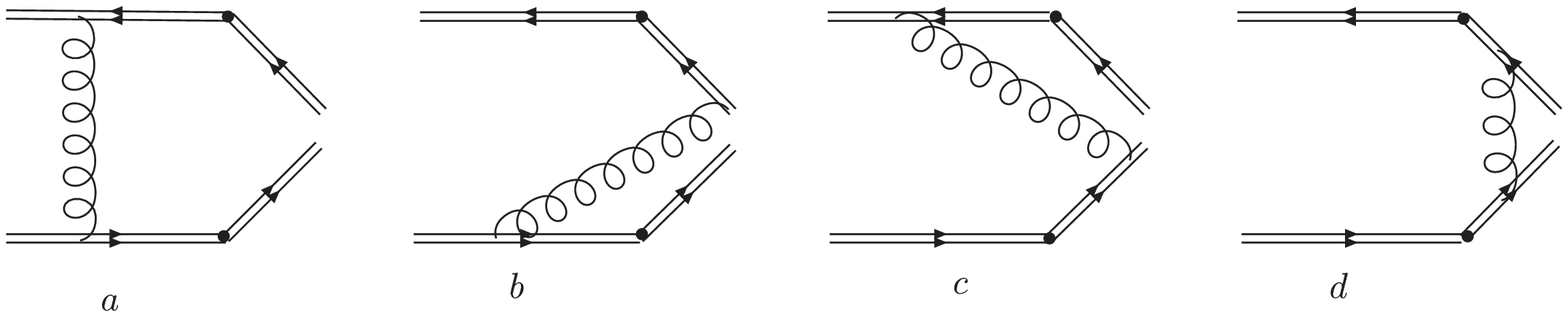}
\end{center}
\caption{The  one-loop contribution to products $S$ of gauge links.}
\label{Feynman-dg1}
\end{figure}
\par
\par
\begin{figure}[hbt]
\begin{center}
\includegraphics[width=7cm]{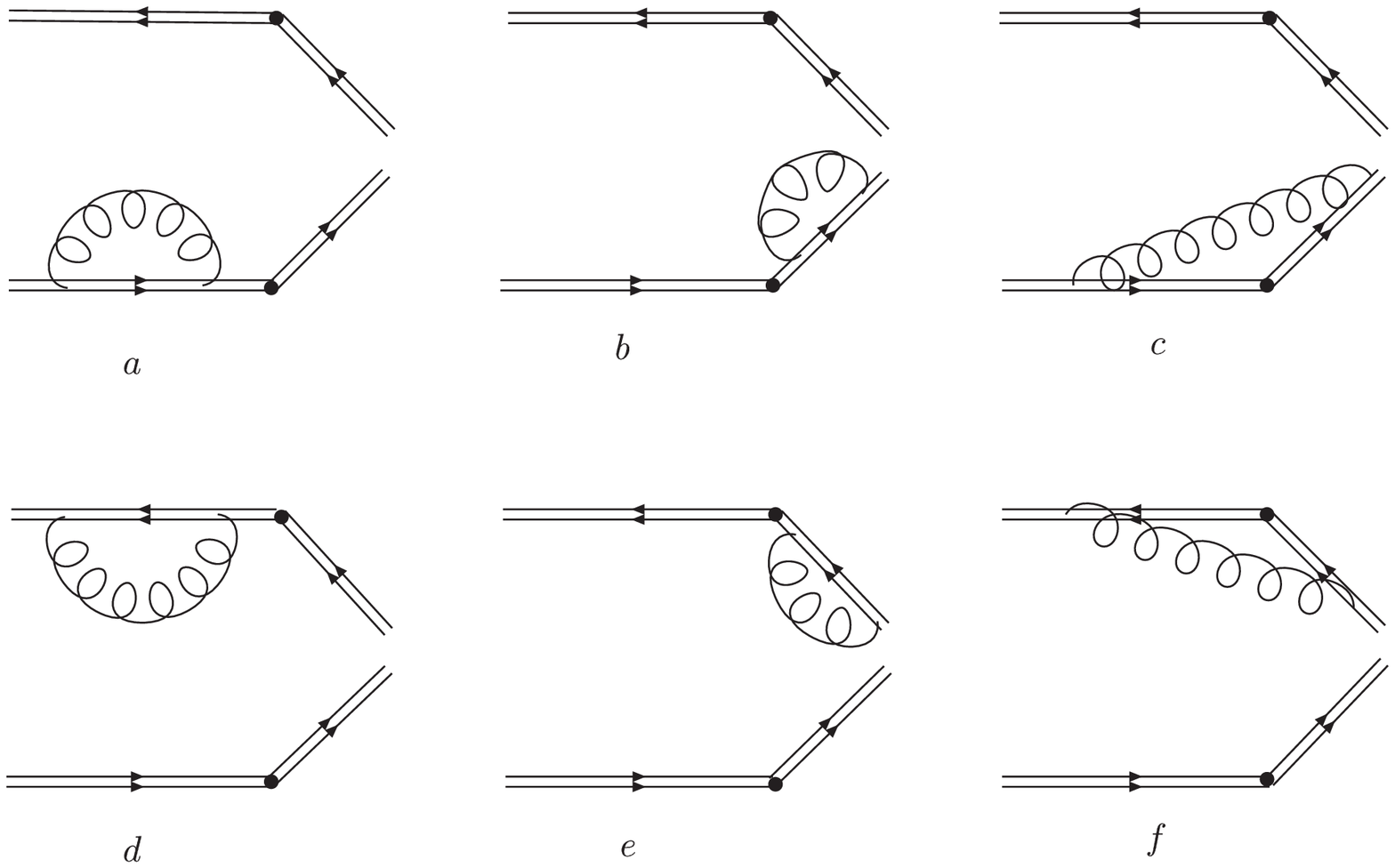}
\end{center}
\caption{The  one-loop contribution to products $S$ of gauge links.}
\label{Feynman-dg1}
\end{figure}
\par
We consider the following products of gauge links:
\begin{eqnarray}
    S(z^-, u,v) &=& \frac{1}{N_c} {\rm Tr } \langle 0 \vert L_v^\dagger (0,-\infty) L_u^\dagger(\infty, 0)
       L_u (\infty, z^-n ) L_v (z^-n, -\infty ) \vert 0 \rangle ,
\nonumber\\
    S(z^-, u,n) &=& \frac{1}{N_c} {\rm Tr } \langle 0 \vert L_n^\dagger (0,-\infty) L_u^\dagger(\infty, 0)
       L_u (\infty, z^-n ) L_n (z^-n, -\infty ) \vert 0 \rangle ,
\nonumber\\
    S(z^-, n,v) &=& \frac{1}{N_c} {\rm Tr } \langle 0 \vert L_v^\dagger (0,-\infty) L_n^\dagger(\infty, 0)
       L_n (\infty, z^-n ) L_v (z^-n, -\infty ) \vert 0 \rangle .
\end{eqnarray}
The vector $v$ is taken as $v^\mu =(v^+, v^-, 0,0)$ with $v^+ >> v^-$.
The fourier transformed $S$ is:
\begin{equation}
S (q^+ , u,v)  = \int \frac{ d z^-}{2\pi} e^{i q^+ z^-}S(z^-, u,v).
\end{equation}
At tree-level all $S$'s are $1$ in the $z^-$-space or $\delta (q^+)$ in the $q^+$-space.
At one-loop level, they receive corrections from Fig.5 and Fig.6.
It is interesting to note that there are certain relations between contributions
of the three gauge link products. E.g., the contribution from Fig.5b to
$S(q^+,u,v)$ is:
\begin{equation}
S(q^+,u,v)\vert_{5b} = i\frac{4}{3} g_s^2   \int \frac{d^4 k}{(2\pi)^4}\delta (k^+ +q^+)
\frac{ u\cdot v} {v\cdot k - i0} \cdot \frac{1}{u\cdot k -i0}\cdot \frac{1}{k^2 -\lambda^2 +i0}.
\end{equation}
It is interesting to note that under the limit $v^+ >> v^-$ and $u^- >> u^+$:
\begin{equation}
\frac{ u\cdot v} {(v\cdot k - i0)(u\cdot k -i0)} \approx \frac{1}{n\cdot k}
\left [ \frac{ n\cdot v}{v\cdot k-i0} -\frac{ n\cdot u}{u\cdot k -i0} \right ],
\end{equation}
where the first term corresponds to the contribution from Fig.5b to $S(q^+,n,v)$,
and the second term corresponds to that to $S(q^+,u,n )$,
Hence we have:
\begin{equation}
 S(q^+,u,v)\vert_{5b} = S(q^+,n,v)\vert_{5b} -S(q^+,u,n)\vert_{5b}.
\end{equation}
The same result also holds for Fig.5c and Fig.6e and Fig.6f. With this observation
we define the soft factor:
\begin{eqnarray}
\tilde S( z^-, \zeta_u ) &=& \frac{1} {2}\left [  1+S(z^-,u,v)-S(z^-,n,v) +S(z^-,u,n) \right ],
\nonumber\\
\tilde S( q^+, \zeta_u ) &=& P^+ \int \frac{ d z^-}{2\pi} e^{i q^+ z^-}\tilde S( z^-, \zeta_u ).
\end{eqnarray}
With above results we have up to one-loop level:
\begin{eqnarray}
\tilde S( (x-x_0) P^+, \zeta_u ) &=& 1 +
\frac{2\alpha_s}{3\pi } \left \{ \left ( \frac{1}{x-x_0} \right )_+
 \left [\theta (x_0-x) -\theta (x-x_0) \right ]
\right.
\nonumber\\
  && \left. \ \ \ \ \ \ \ \ \  -\frac{1}{2}\delta (x-x_0) \left [ \ln\frac{\zeta^2_u (1-x_0)^2}{\mu^2}
           + \ln\frac{\zeta^2_u x_0^2}{\mu^2} \right ] \right \} + {\mathcal O} (\alpha_s^2).
\end{eqnarray}
The soft factor $\tilde S$ only receives contributions from Fig.5d and from the self-interaction of a
gauge link, i.e.,  from Fig.6a, Fig.6b, Fig.6d and Fig.6e.
These contributions are the same of those from Fig.3d, Fig.4b and Fig.4c to the NLCWF, respectively.
Therefore we can use this fact to subtract the contributions to $\phi_+$ from Fig.3d, Fig.4b and Fig.4e.
It should be noted that the soft factor used here may be not unique. This non-uniqueness may be fixed
through a study of higher orders and will not affect our one-loop results
in this work.  Also in the soft factor $\tilde S$ one can take
the limit $v \to l$ so that $\tilde S$ does not depend on $v$.
\par
We modify the definition of the NLCWF as:
\begin{eqnarray}
\tilde  \phi_{+}(x, \zeta, \mu) = \ \int \frac{ d z^- }{2\pi}
    e^{ik^+z^- }
 \frac{ \langle 0 \vert \bar q(0) L_u^\dagger (\infty, 0)
  \gamma^+ \gamma_5 L_u (\infty,z^-n ) q(z^-n ) \vert \pi^0(P) \rangle}
  { \tilde S( z^-, \zeta_u) }.
\end{eqnarray}
Its individual one-loop contributions are
\begin{eqnarray}
\tilde \phi_+(x,\zeta)\vert_{3b}+\phi_+(x,\zeta)\vert_{4c} &=& \frac{2 \alpha_s}{3\pi} \phi_0
\left \{ \theta (x_0-x) \left ( -\frac{ x}{ x_0}\cdot\frac{1}{x-x_0}\right )_+
 \ln \frac{\zeta^2 x_0^2}{m_q^2}
\right.
\nonumber\\
 && \left. \ \ \ \ \ \  + \delta (x-x_0)\left [ \frac{1}{2} \ln \frac{\mu^2}{\zeta^2 x_0^2} -\frac{\pi^2}{6} +1 \right ]
\right \} \phi_0,
\nonumber\\
\tilde\phi_+(x,\zeta)\vert_{3c}+\phi_+(x,\zeta)\vert_{4f} &=& \frac{2 \alpha_s}{3\pi} \phi_0
\left \{ \theta (x -x_0) \left ( \frac{\bar x}{\bar x_0}\cdot\frac{1}{x-x_0}\right )_+
 \ln \frac{\zeta^2 \bar x_0^2}{m_q^2}
\right.
\nonumber\\
 && \left. \ \ \ \ \ \  + \delta (x-x_0)
 \left [ \frac{1}{2} \ln \frac{\mu^2}{\zeta^2 \bar x_0^2} -\frac{\pi^2}{6} +1 \right ]
\right \} \phi_0,
\end{eqnarray}
it should be noted that there is no term like $(\ln (x-x_0)/(x-x_0))_+$ in comparison with $\phi_+(x,\mu)$,
hence it will not lead to any term with $\ln^2 x$ when convoluted with $H^{(0)}$.
The one loop result for $\tilde\phi$ reads:
\begin{eqnarray}
\tilde  \phi_{+}^{(1)}(x, \zeta, \mu) &=&
\frac{2 \alpha_s}{3\pi}
\left \{ \theta (x_0-x) \left ( -\frac{ x}{ x_0}\cdot\frac{1}{x-x_0}\right )_+
 \ln \frac{\zeta^2 x_0^2}{m_q^2} +  \theta (x -x_0) \left ( \frac{\bar x}{\bar x_0}\cdot\frac{1}{x-x_0}\right )_+
 \ln \frac{\zeta^2 \bar x_0^2}{m_q^2}
\right.
\nonumber\\
 && \left.   + \delta (x-x_0)\left [ \frac{1}{2} \ln \frac{\mu^2}{\zeta^2 x_0^2}
        +\frac{1}{2} \ln \frac{\mu^2}{\zeta^2 \bar x_0^2} -\frac{\pi^2}{3} +2 \right ]
\right \} \phi_0 + {\rm Fig.1a}
\nonumber\\
   && + \frac{\alpha_s}{3\pi} \delta (x-x_0)
    \left [ -\ln \frac{\mu^2}{m_q^2} +2 \ln\frac{m_q^2}{\lambda^2} -4 \right ] \phi_0  ,
\end{eqnarray}
the last line is from external-leg corrections.
\par
There is an interesting relation between LCWF and NLCWF. It reads:
\begin{equation}
\tilde \phi_{+} (x,\zeta,\mu) = \int_0^1 dy C(x,y,\zeta,\mu) \phi (y,\mu),
\end{equation}
where the function $C$ can be calculated with perturbative QCD and does not contain any soft
divergence. From our results we have:
\begin{eqnarray}
C(x,y,\zeta, b,\mu) &=& \delta (x-y)+ \frac{2\alpha_s (\mu)}{3\pi} \left \{  \theta(x-y)
\left [\frac{1}{x-y}\left ( \frac{\bar x}{\bar y}\ln\frac{\zeta^2 (x-y)^2}{\mu^2} \right) \right ]_+
\right.
\nonumber\\
 &&  \left. -\theta(y-x) \left [\frac{1}{x-y}
\left ( \frac{x}{y}\ln\frac{\zeta^2 (x-y)^2}{\mu^2}
\right) \right ]_+
\right.
\nonumber\\
&&  \left. + \delta (x-y) \left [ \frac{1}{2} \ln\frac{\mu^2}{\zeta^2 y^2}
+  \frac{1}{2}\ln\frac{\mu^2}{\zeta^2 \bar y^2}
       -\frac{\pi^2}{3} +2 \right ] \right \} + {\mathcal O} (\alpha_s^2).
\end{eqnarray}
With the function we define another function which will be useful later:
\begin{equation}
\frac{ \hat C (x,\zeta,\mu )}{x} =\int_0^1 \frac{dy }{y} C(y,x,\zeta,\mu).
\end{equation}
which is just the convolution of $C$ with the tree-level hard part from Fig.1a. Again the function
has a perturbative expansion:
\begin{eqnarray}
\hat C (x,\zeta,\mu )
   &=& 1 -\frac{2\alpha_s(\mu)}{3\pi} \left \{ \frac{1} {\bar x}
\left [-\ln^2 x + \ln x \ln\frac{\mu^2}{\zeta^2 } + 2 {\rm Li}_2 (x)  -\frac{\pi^2}{3} \right ]
\right.
\nonumber\\
   &&  \left.
    + \frac{1}{2} \ln\frac{\mu^2}{\zeta^2 x^2}
 +  \frac{1}{2}\ln\frac{\mu^2}{\zeta^2 \bar x^2}
        + \frac{\pi^2}{3} +2  \right \} + {\mathcal O} (\alpha_s^2),
\end{eqnarray}

\vskip20pt
\noindent
{\large\bf 4. The Jet Factor and Re-Factorization}
\par\vskip15pt
After having studied the double log in LCWF, we need now to study how to factorize
the double log $\ln^2 x_0$ from the form factor from Fig.2c.
As discussed before,  the double log comes from the loop-momentum region where all
components of the momentum carried by the gluon are at order of $x_0Q$. One can
use the eikonal approximation to expand the contribution in $x_0 $ before the
loop integration to obtain the dominant contribution. After some algebra we have:
\begin{eqnarray}
\langle \gamma (p, \epsilon^*)\vert J^\mu_{e.m.}\vert k_q, k_{\bar q}\rangle \vert_{2c}
  & = &  \left [ -i\frac{4 g_s^2 }{3} \int \frac{ d^4 k }{(2\pi)^4}
  \frac{1}{k^2 +i0} \cdot \frac{1}{(k+k_q-p)^2 -m^2 +i0}\cdot \frac{p^-}{-2 k^- + i0} \right ]
\nonumber\\
  && \cdot \langle \gamma (p, \epsilon^*)\vert J^\mu_{e.m.}\vert k_q, k_{\bar q}\rangle \vert_{1a}
     + {\mathcal O} (x_0^0) ,
\end{eqnarray}
where we omitted irrelevant factors. The eikonal propagator $1/(-2k^- +i0)$ comes
from the quark propagator from the anti-quark after emitting the gluon. This suggests
that we can replace the anti-quark line with a suitable gauge link. However if we take
the gauge link along the light-cone direction, it will produce a light-cone singularity
in the integration over $k^-$ as indicated above. To avoid this we can take
the gauge link along non light-cone direction.
\par
For our purpose we consider the following time-ordered product of gauge links with quark fields:
\begin{eqnarray}
S_q(z) &=&   \frac{1}{6i}{\rm Tr} \left \{ \gamma^- \langle 0 \vert T \left [
V_v^\dagger (0, -\infty )  q (0) \   \bar q(z) V_{\tilde u} (z,-\infty)
 \right ] \vert 0 \rangle \right \},
\nonumber\\
S_q(q) &=& \int d^4 z e^{- i \cdot x} S_q(x),
\end{eqnarray}
with $q^\mu =(xP^+, -p^-,0,0)$. Without the gauge links, it is just a quark propagator.
We first fix the vector $v$ with $v^+ >> v^-$ as discussed above.
The gauge link with $\tilde u$ is needed to make $S_q(q)$ gauge invariant.
The direction of $\tilde u$ will be give later.
At tree-level we have
\begin{equation}
S_q(q) =  \frac{1}{6i}{\rm Tr}\left [ \frac{i\gamma^- (\gamma\cdot q +m)}{ q^2 -m^2} \right ] = \frac{1}{q^+}.
\end{equation}
\par
\par
\begin{figure}[hbt]
\begin{center}
\includegraphics[width=8cm]{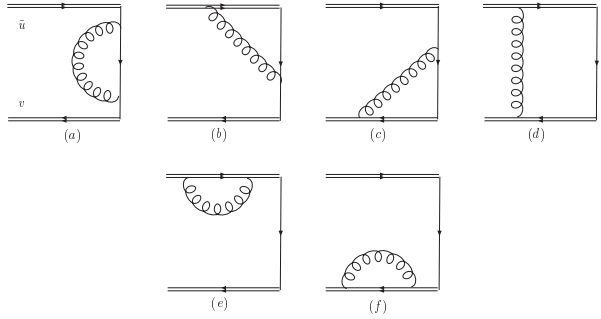}
\end{center}
\caption{One-loop correction of $S_q$.}
\label{Feynman-dg3}
\end{figure}
\par
At one-loop  there are  corrections from diagrams given in Fig.7.
The dominant contribution from Fig.7c
is proportional to the factor in $[ \cdots ]$ in Eq.(38) if the eikonal
propagator $1/(-2k^- +i0)$ is replaced with $1/(-2 v\cdot k +i0)$.
Hence it will produce the same $\ln^2 x$ as that
in the form factor from Fig.2c.
However,
there are contributions involved with  interactions only between gauge links.
They are those from Fig. 7d. Fig 7e and Fig.7f.
If we take the direction $\tilde u$ as  $\tilde u^\mu =(0,0, \tilde u^1, \tilde u^2)$,
the contribution from Fig.7d
can be eliminated
because of $\tilde u \cdot v =0$. The other two can be subtracted with expectation
value of gauge links.
We define the following jet factor as:
\begin{eqnarray}
\hat J(x,\zeta_\gamma, Q, \mu) &=&   \frac{q^+}{6i}\int d^4 z e^{- iq \cdot z}
\frac{  {\rm Tr} \left \{ \gamma^- \langle 0 \vert T \left [
V_v^\dagger (0, -\infty )  q (0) \   \bar q(z) V_{\tilde u} (z,-\infty)
 \right ] \vert 0 \rangle \right \}}
  {{\rm Tr}  \langle 0 \vert T \left [
V_v^\dagger (0, -\infty )  V_{\tilde u} (0,-\infty)
 \right ] \vert 0 \rangle },
\nonumber\\
q^\mu &=& (xP^+, -p^-, 0,0), \ \ \  Q^2 =2P^+ p^-,\ \ \  \zeta_\gamma^2 = \frac{2 v^+ (p^-)^2}{v^-},\ \ \
\end{eqnarray}
with the denominator the contribution from Fig.7e and Fig.7f are subtracted.
The tree-level contribution to $\hat J$ is 1.
The one-loop contributions are then from Fig.7a, Fig.7b and Fig.7c. They
are
\begin{eqnarray}
\hat J(x,\zeta_\gamma, Q, \mu) \vert_{7a} &=&
-\frac{\alpha_s}{3\pi} \left [ \ln\left (\frac{\mu^2}{-q^2} \right ) +1 \right ],
\nonumber\\
\hat J(x,\zeta_\gamma, Q, \mu) \vert_{7b} &=& \frac{\alpha_s}{3\pi q^+ }  \left [
             \ln \left ( \frac{\mu^2}{-q^2} \right ) +2  \right ],
\nonumber\\
\hat J(x,\zeta_\gamma, Q, \mu) \vert_{7c} &=&  \frac{\alpha_s}{3 \pi }  \left [
  \ln \left ( \frac{\mu^2}{-q^2} \right ) +2  + \ln \frac{\zeta^2_\gamma}{xQ^2}
     - \ln^2 \frac{\zeta^2_\gamma}{xQ^2}  -\pi^2 -4 \right ]+ {\mathcal O} (\hat \zeta_\gamma^{-1}).
\end{eqnarray}
At the order we consider $\hat J$ does not depend on $\tilde u^2$. We note that the $\ln^2 x$ term
from Fig. 7c is exactly that from Fig.2c contributing to the form factor as expected.
With the above result one can derive the following evolution equations which
will be useful for our resummation:
\begin{eqnarray}
\frac{\partial}{\partial \ln \mu^2} \hat J(x,\zeta_\gamma, Q, \mu) &=&
 \frac{\alpha_s(\mu)}{3\pi } \hat J(x,\zeta_\gamma, Q, \mu) ,
\nonumber\\
\frac{\partial}{\partial \ln \zeta_\gamma^2} \hat J(x,\zeta_\gamma, Q, \mu) &=&
 \frac{\alpha_s(\mu)}{3\pi }\left ( -2 \ln \frac{\zeta^2_\gamma}{xQ^2} +1 \right ) \hat J(x,\zeta_\gamma, Q, \mu).
\end{eqnarray}
\par
With NLCWF and the jet factor we can write a factorized form for the form factor:
\begin{equation}
  F (Q) \sim \int_0^1 dx \left [ \frac{1}{x} \tilde \phi_+ (x,\zeta,\mu) \hat J(x,\zeta_\gamma, Q, \mu)
  \tilde H (x,\zeta,\zeta_\gamma,Q, \mu) + (x \to 1-x) \right ] ,
\end{equation}
and $\tilde H(x)$ does not contain $\ln^2 x$ explicitly. The leading term $\tilde H$ is one in the above.
With our NLCWF we have the convolutions corresponding to that with LCWF in Eq.(17), subtracted
with the corresponding contributions from the form factor:
\begin{eqnarray}
Q^2 F(Q^2) \vert_{2c} -\int_0^1 \frac{dx}{x}\tilde \phi(x,\zeta) \vert_{3c +4f}  &=&
\frac{2\alpha_s}{3 \pi x_0} \phi_0 \left [ \frac{1}{\bar x_0} \left ( -\frac{1}{2} \ln^2 x_0
+ \ln x_0 \ln\frac{Q^2}{\zeta^2 } +2 {\rm Li}_2 (x_0)
\right.\right.
\nonumber\\
 && \left. \left. -\frac{\pi^2}{3}\right )
 -\frac{1}{2} \ln\frac{\zeta^2 \bar x_0^2}{Q^2}
 -\frac{2+x_0}{2 \bar x_0} \ln x_0
          +\frac{\pi^2}{6} -1 \right ],
\nonumber\\
Q^2 F(Q^2) \vert_{2b} -\int_0^1 \frac{dx}{x} \tilde \phi(x,\zeta) \vert_{3b +4c}
&=&
\frac{2\alpha_s}{3 \pi x_0} \left [ \frac{1}{2} \ln \frac{Q^2}{\zeta^2 x_0} +\frac{\pi^2}{6} -1 \right ],
\end{eqnarray}
it is clearly that all collinear singularities related to the quark mass
are factorized into the NLCWF.
\par
With the jet factor we have:
\begin{eqnarray}
\tilde H (x,\zeta,\zeta_\gamma,Q, \mu) &= & 1 + \frac{2\alpha_s(\mu)}{3\pi}
    \left \{  \frac{1}{\bar x} \left [ \ln x \ln\frac{Q^2}{\zeta^2}
       +\frac{1}{2} \ln^2 \frac{Q^2}{\zeta^2_\gamma} + \ln x \ln\frac{Q^2}{\zeta^2_\gamma} \right ]
\right.
\nonumber\\
   && \left. + \left [ \ln\frac{Q^2}{\zeta^2} + \ln \frac{Q^2}{\mu^2} +\frac{1}{2} \ln\frac{Q^2}{\zeta^2_\gamma} \right ]
          + \frac{1}{\bar x} \left ( -\frac{x \pi^2}{3} +2 {\rm Li}_2 (x) \right ) -2 +\frac{\pi^2}{2}- \ln\bar x
\right.
\nonumber\\
   && \left.  - \frac{3x}{2\bar x} \ln x - \frac{x}{2\bar x} \ln^2 \frac{xQ^2}{\zeta^2_\gamma}
   -\frac{x}{\bar x} \ln  x \left ( \ln\frac{Q^2}{\mu^2} -1 +\frac{1}{2} \ln x \right )
   \right\}
   +{\mathcal O}(\alpha_s^2),
\end{eqnarray}
as expected, for fixed $\zeta,\ \zeta_\gamma$ and $\mu$ there are no terms like $\ln^2x$.
Also there are no terms like $\ln x$ without involving other log's.

\par\vskip20pt
\noindent
{\large\bf 5. Resummation}
\par\vskip15pt
If we take $\tilde \phi_+$ as a nonperturbative object entirely, the resummation is really simple,
in which we chose scales like $\mu$, $\zeta$ and $\zeta_\gamma$ so that there is no
large log's. E.g., we can take those scales and obtain $\hat H$ which contains no large log's:
\begin{eqnarray}
\mu^2 & =& Q^2 = \zeta^2 =\zeta^2_\gamma,
\nonumber\\
\tilde H (x,Q,Q,Q, Q) &= & 1 + \frac{2\alpha_s(\mu)}{3\pi}
    \left \{
           \frac{1}{\bar x} \left ( -\frac{x \pi^2}{3} +2 {\rm Li}_2 (x) \right ) -2 +\frac{\pi^2}{2}
 - \ln \bar x
\right.
\nonumber\\
  && \left. - \frac{3x}{2\bar x} \ln x - \frac{x}{2\bar x} \ln^2 x
   -\frac{x}{\bar x} \ln \bar x \left (  -1 +\frac{1}{2} \ln\bar x \right )
  \right\}
   +{\mathcal O}(\alpha_s^2),
\end{eqnarray}
and for the jet factor we use the evolution equation of $\zeta_\gamma$ to express
$\hat J$ at $\zeta_\gamma = Q$ with that at $\zeta_\gamma = \sqrt{x} Q$:
\begin{eqnarray}
\hat J(x,Q, Q, Q) &=& \exp \left \{ -\frac{\alpha_s(Q)}{3\pi}
   \left [ \ln^2 x
                     + \ln x \right ] \right \}
\hat J(x,\sqrt{x}Q, Q, Q),
\nonumber\\
\hat J(x,\sqrt{x}Q, Q, Q) &=&  1 +\frac{\alpha_s}{3\pi} \left [
   -\ln x   -1 -\pi^2 \right ],
\end{eqnarray}
it should be noted that $\hat J(x,\sqrt{x}Q, Q, Q)$ has no $\ln^2 x$ term.
Only the single log $\ln x$ remains. Then for the form factor
we have:
\begin{eqnarray}
  F (Q) & \sim & \int_0^1 dx \left [ \frac{1}{x} \tilde \phi_+ (x,Q) \hat J(x,\sqrt{x}Q, Q, Q)
  \tilde H (x,Q,Q,Q,Q) \exp \left \{ -\frac{\alpha_s(Q)}{3\pi}
   \left [ \ln^2 x
                     + \ln x \right ] \right \}
\right.
\nonumber\\
       && \left.  + (x \to 1-x) \right ] .
\end{eqnarray}
Taking $\tilde H \hat J$ as a perturbative function, it does not contain $\ln^2x$. The terms
with $\ln^2 x$ are resummedin the exponential, but one needs the information of
$\tilde \phi$ to make predictions.
\par
With our results one can indeed resum $\ln^2 x$
in the factorization formula with LCWF. One can use the relation between
$\phi$ and $\tilde\phi$ to write another factorization formula for the form factor:
\begin{equation}
  F (Q) \sim \int_0^1 dx \left [ \frac{1}{x} \phi (x,\mu)
   \hat C (x,\zeta,\mu) \hat J(x,\zeta_\gamma, Q, \mu)
  \hat H (x,\zeta,\zeta_\gamma,Q, \mu) + (x \to 1-x) \right ] ,
\end{equation}
and take $\hat C$,  $\hat J$ and  $\hat H$ as perturbative functions.
$\tilde H$ is the same as $\hat H$ at one-loop level.
If we expand the product $\hat C \hat J \tilde H$, we return to the
standard collinear factorization discussed in Sec.2.
With the product its each part has a clear meaning. The $\zeta$-dependence
in $\hat C$ will control the behavior of $x\to 0$ in LCWF, while the
$\zeta_\gamma$ dependence in $\hat J$ controls that in the form factor.
The evolution equations of $\hat C$ can be obtained from results in Sec.3. They are:
\begin{eqnarray}
\frac{\partial}{\partial \ln \mu^2} \hat C (x,\zeta,\mu) &=& -\frac{2 \alpha_s(\mu)}{3\pi}
     \left ( \frac{1}{\bar x} \ln x +1 \right ) \hat C (x,\zeta,\mu),
\nonumber\\
\frac{\partial}{\partial \ln \zeta^2} \hat C (x,\zeta,\mu) &=& \frac{2 \alpha_s(\mu)}{3\pi}
     \left ( \frac{1}{\bar x} \ln x +1 \right ) \hat C (x,\zeta,\mu).
\end{eqnarray}
\par
For the resummation we first chose a scale $\mu_1$ in the factorization formula
and use the $\mu$-evolution
to express $\hat C(x,\zeta,\mu_1)$ with $\hat C(x,\zeta,\mu)$:
\begin{equation}
\hat C(x,\zeta,\mu_1) = \exp \left \{ - \frac{8}{3\beta_0} \ln \frac{\alpha_s(\mu)} {\alpha_s(\mu_1)}
\left (\frac{\ln x}{\bar x} +1 \right ) \right \}\hat C(x,\zeta,\mu),
\end{equation}
where we have used the one-loop $\alpha_s$-running:
\begin{equation}
  \alpha_s(\mu) = \frac{4\pi}{\beta_0 } \left (\ln\frac{\mu^2}{\Lambda^2} \right )^{-1}, \ \ \ \
  \beta_0 = 11-\frac{2}{3} n_f.
\end{equation}
We then use the $\zeta$-evolution to express $\hat C(x,\zeta,\mu)$ with $\hat C(x,\zeta_0,\mu)$:
\begin{equation}
\hat C(x,\zeta,\mu) = \exp \left\{ \frac{2\alpha_s(\mu)}{3\pi} \left ( \frac{\ln x}{\bar x} +1 \right )
   \ln \frac{\zeta^2}{\zeta^2_0} \right \}\hat C(x,\zeta_0,\mu),
\end{equation}
now we take $\zeta_0^2 = \mu^2 /x$, so that $\hat C(x,\zeta_0,\mu)$ does not have any log's:
\begin{equation}
\hat C(x,  \mu/\sqrt{x},\mu) = 1 -\frac{\alpha_s (\mu)}{3\pi} \left [ \frac{1}{\bar x}
\left ( 2 {\rm Li}_2 (x) -\frac{\pi^2}{3} \right ) -\frac{1}{2}\ln \bar x +\frac{\pi^2}{3} +2 \right ]
   + {\mathcal O}(\alpha_s^2).
\end{equation}
With these steps, all log terms in $\hat C$ are resumed:
\begin{equation}
\hat C(x,\zeta,\mu_1) = \exp \left \{ - \left [\frac{8}{3\beta_0} \ln \frac{\alpha_s(\mu)} {\alpha_s(\mu_1)}
 -\frac{2\alpha_s(\mu)}{3\pi} \ln \frac{x \zeta^2}{\mu^2} \right]
  \left ( \frac{\ln x}{\bar x} +1 \right ) \right \} \hat C(x,  \mu/\sqrt{x},\mu).
\end{equation}
Now we have the freedom to chose $\mu$ so that the exponential does not go to $\infty$ when $x$ goes to $0$.
We can take $\mu$ fixed by:
\begin{equation}
  \alpha_s (\mu)  = x \alpha_s (\mu_1), \ \ \ \  \ln\frac{\mu^2}{\Lambda^2} = \frac{1}{x} \ln\frac{\mu_1^2}{\Lambda^2},
\end{equation}
and
\begin{equation}
\hat C(x,\zeta,\mu_1) = \exp \left \{ - \frac{8}{3\beta_0}\left [ \ln x
  -\frac{\beta_0}{4\pi} x \alpha_s(\mu_1) \ln x -x \frac{\alpha_s(\mu_1)}{\alpha_s (\zeta)} +1 \right]
  \left ( \frac{\ln x}{\bar x} +1 \right ) \right \} \hat C(x,  \mu/\sqrt{x},\mu).
\end{equation}
For $x\to 0$ we have now:
\begin{equation}
\hat C(x,\zeta,\mu_1) \sim \exp \left \{ - \frac{8}{3\beta_0}\ln^2 x \right \},
\end{equation}
and it goes to zero fast than any positive power of $x$.
\par
By taking the scales
\begin{equation}
\mu_1^2 =Q^2 =\zeta^2 =\zeta^2_\gamma,
\end{equation}
and using the $\zeta_\gamma$-evolution to express $\hat J$ at $\zeta^2_\gamma =Q^2$
with $\hat J$ at $\zeta^2_\gamma =xQ^2$, we obtain our resummedform for the form factor:
\begin{eqnarray}
  F (Q) & \sim & \int_0^1 dx \left [ \frac{1}{x} \phi (x,Q) \hat J(x,\sqrt{x}Q, Q, Q)
  \hat H (x,Q,Q,Q,Q)  \hat C(x,  \mu/\sqrt{x},\mu)  \exp \left \{ -S(x,Q)\right \}
\right.
\nonumber\\
  && \left. + (x \to 1-x) \right ] ,
\nonumber\\
   S(x, Q) &=& \frac{8}{3\beta_0}\left [ \ln x
  -\frac{\beta_0}{4\pi} x \alpha_s(Q) \ln x   -x +1 \right]
  \left ( \frac{\ln x}{\bar x} +1 \right )+ \frac{\alpha_s(Q)}{3\pi}\left [ \ln^2 x +\ln x \right ],
\nonumber\\
          &\approx &  \left ( \frac{8}{3\beta_0} + \frac{\alpha_s(Q)}{3\pi} \right ) \ln^2 x, \ {\rm for} \
           x \to 0,
\end{eqnarray}
in the above the product $\hat C \hat H \hat J$ does not contain any log's, except $\hat J$
has a single log $\ln x$. All other logs, like $\ln^2 x$,etc., are resummedin $S$. Since we only used
one-loop evolutions, for consistence we should neglect higher orders in $\alpha_s$ in the
product. Therefore, we have the one-loop resummedform factor:
\begin{eqnarray}
  F (Q) &= & \frac { 1}{3 \sqrt{2} Q^2}
   \int_0^1 dx \left [ \frac{1}{x} \phi (x,Q)  \exp \left \{ -S(x,Q)\right \}
           + (x \to 1-x) \right ] ,
\nonumber\\
     &=& \frac{ \sqrt{2} }{ 3 Q^2}
      \int_0^1 dx  \frac{1}{x} \phi (x,Q)  \exp \left \{ -S(x,Q)\right \} .
\end{eqnarray}
This form can be used if one can get $\phi (x,Q)$ easily at the large scale $Q$. If one only knows
the wave function at lower scale, but does not want to solve the evolution equation for the
wave function to get it at a higher scale, one can first evolute everything at a lower scale
$\mu_0$ where the wave function is known or modeled, then to a higher scale $\mu_1$ and perform
the resummation.
We get in this case:
\begin{eqnarray}
F(Q)= \frac{ \sqrt{2} }{ 3 Q^2}
      \int_0^1 dx  \frac{1}{x} \phi (x,\mu_0)  \exp \left \{ -S(x,Q)
     -\frac{8}{3 \beta_0}\ln\frac{\alpha_s (Q)}{\alpha_s(\mu_0)}
  \left ( \ln x +\frac{3}{2} \right )\right \},
\end{eqnarray}
where we have used:
\begin{equation}
   \frac {\partial \left (  \hat C \hat H \hat J \right )}{\partial \ln\mu^2} = -\frac{2\alpha_s(\mu)}{3\pi}
   \left ( \ln x +\frac{3}{2} \right ) \left ( \hat C \hat H \hat J \right ).
\end{equation}
\par\vskip20pt
\noindent
{\large\bf 6. Numerical Results and Comparison with Experiment}
\par\vskip15pt
We will use our resummation formula in Eq.(62) and Eq.(63) to give our numerical results.
In our formulas the nonperturbative input is the LCWF. The LCWF has the asymptotic form
if $\mu$ goes to $\infty$:
\begin{equation}
  \phi (x, \mu) = 6 x(1-x) f_\pi + \cdots,
\end{equation}
where $\cdots$ stand for terms which are zero in the limit $\mu \to \infty$.
The LCWF can be expanded with Gegenbauer
polynomials\cite{BL}.
A model for $\phi$ has been proposed by truncating the expansion\cite{BrFi}:
\begin{equation}
\phi (x, \mu) = 6 f_\pi x(1-x ) \left ( 1
+  \phi_2 (\mu) C_2^{3/2}(2x-1) \right ) ,
\end{equation}
where $\phi_2 (\mu_0)$ is determined by QCD sum-rule method at $\mu_0 =1$GeV\cite{BrFi}:
\begin{equation}
\phi_2 (\mu_0=1{\rm GeV}) =0.44.
\end{equation}
We will use these two types of LCWF to give our numerical results.
We will use Eq.(61) with the asymptotic form of $\phi$ to make our numerical
predictions. For LCWF given in Eq.(65) we use Eq.(62). We take the $\Lambda$-parameter
as $\Lambda=237$MeV. Our numerical results do not strongly depend on the value of $\Lambda$.
There is only a little change if we change $\Lambda$ from $100$MeV to $300$MeV.
\par

\begin{figure}[hbt]
\begin{center}
\includegraphics[width=14cm]{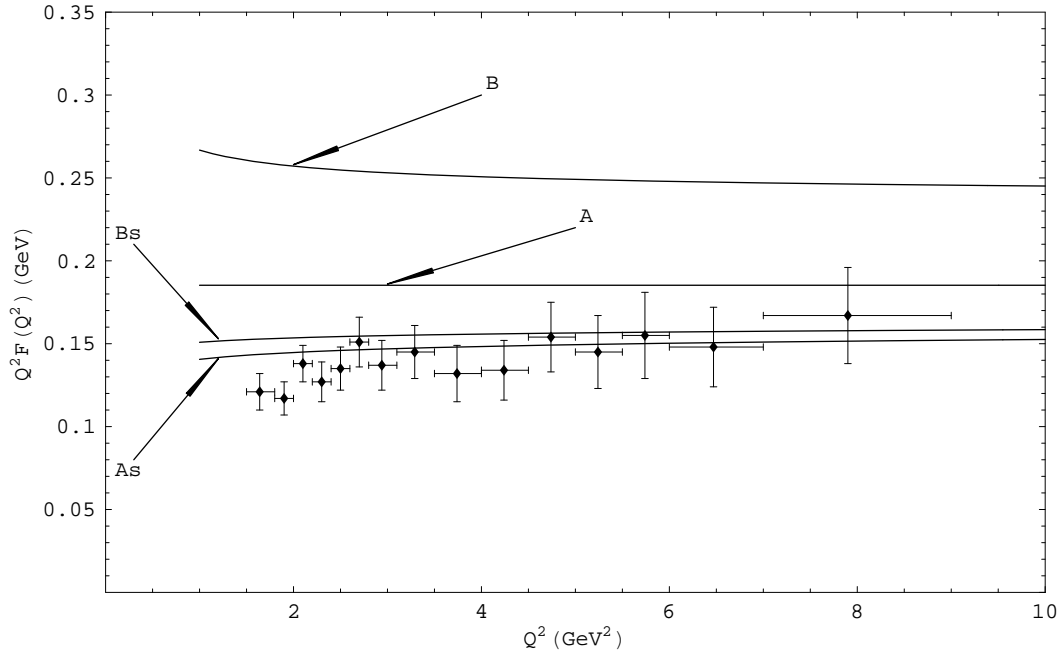}
\end{center}
\caption{Numerical results with experimental data. The curve A and curve As are obtained
by using the asymptotic form without and with the resummation, respectively. The curve B and
curve Bs are obtained
by using the LCWF in Eq.(65) without and with the resummation, respectively.
The experimental data are taken from the second reference in \cite{EXP}.   }
\label{Feynman-dg8}
\end{figure}
\par
Our numerical results are given in Fig.8, where experimental results from CLEO in \cite{EXP}
are also given. From Fig.8 we see that with the two types of LCWF
the resummation has significant effects. The resummation can reduce the form factor predicted
without the resummation at the level of $40\%$ or more. Using the LCWF given in Eq.(65) with our resummation
the experimental results can be well described for $Q^2\geq 3{\rm GeV}^2$.

\par\vskip20pt
\noindent
{\large\bf 7. Conclusion}
\par\vskip15pt
In the collinear factorization the form factor of the transition
$\gamma^* \pi^0 \to \gamma$ can be written as
a convolution of a hard part and LCWF.
The hard part contains double log terms as $\ln^2 x$ at one-loop level
and is expected to have terms $\ln^{2n} x$ at order of $\alpha_s^n$.
A resummation of these terms with
a simple exponentiation can not be done because it results in divergent
results.
In this work we have studied the resummation of these $\ln^2 x$ terms.
With a small but finite quark mass as the regulator of collinear singularities,
we have found that the $\ln^2 x$ terms come partly from the light-cone wave function
and partly from the form factor, as discussed in Sec.2. To handel
these terms, we first introduce a nonstandard light-cone wave function with
the gauge links off the light-cone direction.
This introduces an extra scale in the NLCWF beside the renormalization scale $\mu$.
The deviation from the light-cone
direction will regularize light-cone singularities in each contributions.
This fact leads to that the NLCWF will not deliver any term with $\ln^2 x$
to the hard part, if one uses the NLCWF to perform the factorization.
As the next, we introduce a jet factor to factorize the $\ln^2 x$ term in the form factor.
The jet factor also contains an extra scale beside $\mu$. This extra scale
controls the $x$-behavior of the jet factor.
Our re-factorized formula for the form factor is a convolution with the NLCWF, the jet factor
and a hard part. The hard part does not contain terms with $\ln^2 x$.
\par
We have found that there is an interesting relation
between the LCWF and the introduced NLCWF. The relation
can be determined with perturbative QCD and is given at one-loop level in this work.
With this relation we are able to show that the $\ln^2 x$ can be resumed
and the nonperturbative object in the resummed formula is
only the LCWF. With the knowledge of LCWF's we are able to get numerical predictions.
In performing the resummation of the double log we have used
the concept of QCD factorization and worked out every quantity explicitly
at one-loop level. It is possible to extend our work to the resummation
of the remaining single log terms and beyond one-loop level.
\par
Our numerical results show that the effect of the resummation is significant.
There is a difference at the level of $40\%$  or more between the predicted form factors
with and without the resummation. In comparison with experiment we find that the numerical
predictions by using the LCWF in Eq.(65) with the resummation are consistent with the experimental
data.

\par\vskip20pt
\noindent
{\bf\large Acknowledgments}
\par
This work is supported by National Nature Science Foundation of P.R.
China(No. 10575126, No. 10421003).
\par


\end{document}